*Article*

# Towards a prioritised use of transportation infrastructures: the case of vehicle-specific dynamic access restrictions to city centres

Holger Billhardt [1], Alberto Fernández [1,*], Pasqual Martí [2], Javier Prieto Tejedor [3] and Sascha Ossowski [1]

[1] CETINIA, Univ. Rey Juan Carlos, 28933 Móstoles, Madrid, Spain; holger.billhardt@urjc.es (H.B.); sascha.ossowski@urjc.es (S.O.)
[2] Valencian Research Institute for Artificial Intelligence (VRAIN), Universitat Politècnica de València, Spain; pasmargi@vrain.upv.es
[3] BISITE Research Group, University of Salamanca. Edificio Multiusos I+D+i, 37007, Salamanca, Spain; javierp@usal.es
* Correspondence: alberto.fernandez@urjc.es

**Abstract:** One of the main problems that local authorities of large cities have to face is the regulation of urban mobility. They need to provide the means to allow efficient movement of people and distribution of goods. However, the provisioning of transportation services should take into account general global objectives, like reducing emissions and having more healthy living environments, which may not always be aligned with individual interests. Urban mobility is usually provided through a transport infrastructure that includes all the elements that support mobility. In many occasions, the capacity of using elements of this infrastructure is lower than the actual demand and thus, different transportation activities compete for using them. In this paper, we argue that scarce transport infrastructure elements should be assigned dynamically and in a prioritised manner to transport activities that have a higher utility from the point of view of the society, for example, activities that produce less pollution and provide more value to society. In this paper, we define a general model for prioritizing the use of a particular type of transportation infrastructure elements, called time-unlimited elements, whose usage time is unknown a priori, and illustrate its dynamics through two use cases: vehicle-specific dynamic access restriction to city centres i) based on the usage levels of available parking spaces and ii) to assure sustained admissible air quality levels in the centre. We carry out several experiments using the SUMO traffic simulation tool, to evaluate our proposal.

**Keywords:** traffic management; prioritized resource allocation; urban mobility; agreement technologies

## 1. Introduction

The organization of urban mobility and transportation is a field that has received tremendous changes, as well as a remarkable interest in the last years. Not only the desire of people to move freely within the cities, but also the transformation of customer habits towards an increasingly online acquisition of goods and the subsequent logistic requirements, tend to increase the traffic in big cities. At the same time the consciousness and sensibility has grown regarding environmental pollution and its effects on public health and the quality of life of citizens. In this context, authorities of big cities all over the world are faced with the problem of providing efficient transportation solutions reducing at the same time traffic-related problems like traffic jams or environmental pollution. In parallel to this trend, both research and industry, have proposed and provided new innovative solutions for more environment-friendly means of transportations. Examples of this trend are new types of vehicles (e.g., electric cars, scooters, bikes and so on) or new transportation services that are based on the concept of "collaborative economy" or "collaborative consumption" [1] and aim at a more efficient usage of available resources, e.g., the sharing of transportation means for several transportation tasks.

We argue that in this new context also the management of the usage of transportation infrastructures can help to control environmental pollution while facilitating at the same time the transportation of goods and people in big cities. Here we understand as infrastructures, all those facilitating elements or resources of a transport system that are used in a shared manner by different users and at different times, such as intersections, roads, tracks and lanes, parking spaces, etc. The availability of such resources is usually limited, and we believe that new usage schemes should be set up that prioritize vehicles or transportation services that are more environment-friendly and have a higher transportation efficiency or are more important from a social point of view. A simple, already existing example of this idea is the ability of ambulances to cross any crossroad when they carry patients in life-threatening situations.

In this paper, we set out a general model for assigning limited traffic infrastructure resources to transportation tasks in a prioritized manner such that a predefined overall utility for the society can be increased. This utility can reflect the cost (pollution, energy, consumption, etc.), as well as the importance of a trip from the point of view of the society. Our approach concentrates on, what we call, *time-unlimited infrastructure elements*, elements that are assigned for an a priori unknown usage time. In this case, the assignment is often accomplished in a one-shot manner without temporal planning in the short term. Examples of such elements are parking spaces, access to restricted areas or roads, vehicles from sharing systems, etc. In the second part of the paper, we instantiate the proposed model in two concrete use cases: vehicle-specific dynamic access restriction to city centres i) based on the usage levels of available parking spaces and ii) to assure sustained admissible air quality levels in the centre. For both use cases, we analyse the trade-off between quality of service and emission levels of various vehicle-specific control strategies, by means of simulation experiments using the SUMO microscopic traffic simulator [2].

The paper is structured as follows. Section 2 presents some related work on prioritizing the use of traffic infrastructures as well as on smart parking management and pollution control in urban areas. Section 3 gives a general formalisation for our prioritized access model. In section 4 we instantiate the model with a use case of parking space assignment in a city centre and present and discuss the results obtained in simulation experiments. In section 5 we study the second use case, the adaptive access control to a restricted area based on its air quality level. Again, we present and discuss an empirical evaluation through simulation experiments. Finally, section 6 concludes the paper.

## 2. Related work

A lot of works can be found in the Intelligent Transportation Systems literature that aim to find smart solutions to traffic control in big cities. The expansion of smart road infrastructures, supported by vehicle-to-vehicle and vehicle-to-infrastructure communications, opened a field for experimenting with a variety of methods and approaches to address different challenges.

In particular, we aim at regulating a prioritised usage of public transportation infrastructures such as traffic lights, road lanes, restricted areas, etc. In this line, [3] proposed the concept of dynamic road space allocation with the goal of optimizing the use of underutilized spaces. The authors pointed out that existing approaches are usually static, thus not flexible enough to account for urban dynamics. They discussed challenges and proposed a methodology for implementing such systems.

The prioritised use of road traffic infrastructures has been commonly regulated by means of traffic lights or smart intersections [4]. While traffic light actions (phases) make no distinction of the vehicles demanding access, intersection management can deal with individual agents. This is the case of the reservation-based control system proposed by [5], which allows autonomous vehicles to negotiate with intersection managers time and space slots to cross the intersection. That system has been extended by different authors, for example by market-based mechanisms to prioritise the access to networks of intersections [6]. Rather than using auctions, like most intersection control approaches, [7] proposed a market-based cooperative framework where vehicles can directly trade their crossing turn based on their "value of time".

The prioritized use of road lanes is another problem that urban managers have to deal with. [8] tackled the problem of deciding which roads can cede a lane for public transportation (e.g. buses). [9] coordinated the operation of buses that use exclusive bus corridors that have bidirectional lanes. Their goal is to minimize the total waiting time of passengers. Dynamic exclusive bus lanes (also known as Bus Lanes with Intermittent Priority) aim to increase the use of bus lanes by allowing other vehicles to use bus lanes when there are no buses using them [10, 11].

We instantiate the idea of prioritised usage of transportation infrastructures to regulate the access to city centres: i) based on available parking spaces and ii) to assure sustained air quality levels in the centre. Our objective is proposing a prioritised access control approach that is highly dynamic, specific to individual vehicles and that considers social utility or transportation efficiency. In this sense, our work falls in the context of Urban Vehicle Access Regulations (UVAR), which can be defined as "measures to regulate vehicular access to urban infrastructure" [12]. UVARs are mostly designed to regulate freight transport, and are based on parameters such as access time, vehicle characteristics (e.g. emission level, size), and load factors [13].

Smart parking systems [14] aim at providing efficient solutions to the parking problem in populated urban areas, which causes an increase of traffic congestion and therefore $CO_2$ emissions, noise, time spent by users, etc. Some works focus on helping users to find available parking places ([15–17]). Others are oriented to balance demand and parking availability. In this sense, dynamic pricing is the most common approach to manage parking occupancy [14]. Prices are usually decided based on parking availability and demand ([18–22]), or after multiagent negotiation processes ([23–25]). In our parking management use case, we approach the problem at a different level. We do not focus on efficient management of specific parking places in an area. Instead, we deal with the problem at a higher level by considering the whole city centre as a virtual parking with the total number of places. Then, we prioritise the reservation of a parking place in the city centre based on vehicle utility.

Traffic control systems are often focused on regulating traffic flow to avoid congestion, thus indirectly reducing pollution. Recently, approaches that specifically account for pollution emissions are gaining interest. [26] analysed two control actions to reduce air pollution in urban areas caused by traffic, namely reducing speed and environmental restricted zone. The latter imposes access restrictions on the most contaminant vehicles based on their classification according to the European Emission Standards. The implementation was static, i.e. vehicles were classified in four categories and the two less polluting categories were permitted access. [27] evaluated different intersection control algorithms, showing that platoon-based algorithms obtain less pollutant emissions (higher throughput) but lower fairness than FIFO. [28] studied the effectiveness of traffic signal control and variable message signs for reducing traffic congestion and pollutant emissions. [29] proposed a Pareto-optimal Max Flow algorithm, which obtains multiple distinct possible paths with maximum flow between a pair of points. Thus, these solutions can be used to distribute traffic and pollution more evenly through a city. These works focus on simulating and assessing the performance of specific static actions. A dynamic traffic light control system based on traffic and air pollution was presented in [30].

[31] highlighted that tyres are an important source of particulate matter (PM), so electric vehicles are not entirely emissions-free. They proposed a distributed access control mechanism to regulate PM generation and fair access to a city zone. Their approach encourages ride sharing by (1) matchmaking cars and passengers, and (2) cars that may enter the city are chosen randomly based on their occupancy by a method that ensures fairness and privacy.

## 3. Prioritised Access to Transport Infrastructures

In this section we propose a general model for a prioritized allocation of *time-unlimited* transport infrastructure resources. As pointed out earlier, we understand by transport infrastructure all elements that are provided to the general public facilitating mobility. Such elements may be static, like for instance, streets, lanes of a street, crossroads, parking spaces, etc. Other elements may be mobile, like vehicles of sharing systems or more classical public transportation facilities, like buses, trains or subways, and so on. Infrastructure elements may be used by any person, and their usage is usually regulated through specific norms or conventions. For example, the usage of a lane of a particular street is regulated through the corresponding traffic norms. Also, many elements may be used without charge (usually this holds for most static elements) and others may have some cost (e.g., public transportation).

Transportation infrastructure elements are intrinsically limited and typical traffic problems like traffic jams or excessive delays in movements arise when the usage demand of certain elements exceeds the available resources or capacities. Such mismatches between demands and available resources usually arise in big cities where the population density is very high. In addition, in many big cities, there might also be an interest in putting additional limitations on the use of certain infrastructure elements. For instance, in many European cities, traffic is restricted in some ways in the centre with the aim of having more human-friendly environments or reducing pollution.

Typically, the aforementioned situations lead to a problem of assigning limited resources to an excessive demand and the decision who should be allowed to access a given resource. In the traffic domain, such decisions are typically not taken in a conscious way. Rather the rule of "who comes earlier wins" applies. In contrast, we belief that from the point of view of improving the social welfare, limited infrastructure capacity should be preferably assigned to users or tasks that are more "important" or less harmful with respect to some global, social parameters. That is, the access to or use of limited transport infrastructures should be prioritized.

In this paper we concentrate on *time-unlimited* infrastructure elements. These are resources that are used by vehicles for an a priori unknown period of time, i.e., the assignment procedure cannot predict how long a vehicle will use the element and the assignment is done until that vehicle releases the resource. Examples of such elements are parking slots, limited access areas, vehicles in sharing systems, etc. In contrast, time-limited resources are those that are used by vehicles for a usually very short time period and their assignment is normally more related to allocate concrete and limited time slots to different requests, e.g. the usage of road lanes, or passing cross roads.

Our model is based on the notion of a *transportation task*. A transportation task refers to the mission to carry a *transportation element* (e.g., person, parcel, etc.) from an origin location to some destination. A transportation task may have some additional constraints, e.g., time constraints or other specific maintenance requirements due to the characteristics of the transportation element. Transportation tasks are accomplished through *trip*s using *vehicles*. Vehicles have different characteristics, like size, type, emissions, etc. A *trip* represents the single movement activity carried out by an individual vehicle for accomplishing one or more transportation tasks.

Each trip will have a certain utility for the user (usually the driver or the person that issued the trip). However, here we are not interested in the utility of the trip for an individual user, but in its utility from the point of view of the whole society (called *global utility*). We define the global utility of a trip $t$ as a function of three parameters as follows:

$$U(t) = g\big(Im(t), QoS(t), C(t)\big). \tag{1}$$

$Im(t)$ represents the importance of the trip for the society and it can be defined as an aggregation of the importance of the different transportation tasks that are carried out with the trip. Let trip $t$ contain the set of transportation tasks $\{tt_1, tt_2, \ldots, tt_n\}$, then $Im(t) = \sum_{i=1}^{n} Im(tt_i)$.

It should be noted that the (social) importance of a task may be aligned with the importance it has for the individual, but this might not be the case in all situations (for instance, an urgent medical transportation versus a person going shopping).

However, in general, the aim of a mobility infrastructure is to provide mobility to the citizens and, thus, implicitly any transportation task has a priori a certain positive importance.

$QoS(t)$ is the expected quality of service of trip $t$. This parameter captures the idea that the utility of a trip depends on how well the transportation tasks are carried out. One of the main factors QoS will include is the duration of a trip. It will typically depend on the traffic situation and the transport infrastructure elements that are available for the underlying trip.

$C(t)$ refers to the cost a trip generates from the point of view of the society (not the cost for the user). This may include direct or indirect costs of the use of the infrastructure or energy resources, produced emissions, fuel consumption, etc.

In general, the global utility of a trip is directly correlated with the importance and the quality of service, and it is inversely correlated with the cost. Thus, one possible way to define $U(t)$ could be:

$$U(t) = \frac{Im(t) \cdot QoS(t)}{C(t)}. \tag{2}$$

However, other functions may be used.

During a trip, a vehicle may want to use time-unlimited elements of the transport infrastructure. As we propose in this paper, the use or assignment of such infrastructure elements should be prioritized in order to optimize global utility. This could be done in the following way. Let $I$ be a time-unlimited infrastructure element with capacity $cap(I)$. Furthermore, let $T = \{t_1, t_2, \ldots, t_m\}$ be the trips that request some of $I$'s capacity in a given instant where $capR(I, t_i)$ denotes the portion of $I$ that is requested by the trip $t_i$. The resources or capacities of $I$ in any instant should be assigned by some *control strategy* whose objective is to assign the resources of $I$ to a subset $T' \subseteq T$ such that the following function is maximized:

$$\sum_{t_i \in T'} U(t_i|I) + \sum_{t_i \in T \wedge t_i \notin T'} U(t_i|\bar{I}) \tag{3}$$

and subject to:

$$\sum_{t_i \in T'} capR(I, t_i) \leq cap(I) \tag{4}$$

The utility of a trip is affected by the assignment of the requested resources. In this sense, $U(t_i|I)$ and $U(t_i|\bar{I})$ denote the expected utility of trip $t_i$ if the requested capacity of $I$ is assigned or is not assigned to $t_i$, respectively. Usually, $U(t_i|I) \geq U(t_i|\bar{I})$ because the quality of service of the different transportation tasks included in trip $t_i$ will be different whether the requested resource is assigned to the trip or not. Here we assume that the vehicle requests the use of the element $I$ in order to provide the best possible quality of service. Thus, if the requested capacity is denied, the vehicle needs to find an alternative solution, which will usually result in a lower quality of service, and thus, decreases the global utility. Normally, the factor that will be directly affected by the denegation of a requested resource would be the duration of the trip.

Given a set $T$ of requests, an optimal decision-making method for the control strategy is given if the resources are assigned to trips $t \in T$ in decreasing order of $U(t_i|I) - U(t_i|\bar{I})$ and up to the maximal capacity of the underlying infrastructure element in the given instance. However, $U(t_i|I)$ and $U(t_i|\bar{I})$ are usually unknown at decision-making time and can only be estimated by $U^*(t_i|I)$ and $U^*(t_i|\bar{I})$, respectively. In particular, the effect of assigning or denying a requested resource on the quality of service may have to be estimated.

In general, we consider that different elements of the transportation infrastructure should be regulated with different control strategies. The general objective here is to increase the global utility of the whole transportation system in a city in terms of the aggregation of the utilities of all transport trips, and this can be obtained by prioritizing the trips with higher utility. In addition, giving privileges to more efficient trips (trips with less cost and higher importance) will promote such trips and may encourage users to invest in vehicles with less social costs or to optimize the loads of their trips.

## 4. Use Case: city centre access restriction due to parking limitations

In this use case, we consider people who want to use their car to enter the centre of a city in order to accomplish some tasks and the number of parking spaces are limited. Thus, the infrastructure element whose usage is regulated through a control strategy is the assignment of the available parking spaces. In this context, we assume that cars that approach the centre of a city try to "reserve" a parking slot and only those cars that get a parking slot are allowed to enter the city centre. Other vehicles have to find parking space outside the central area, and their drivers may take, for example, public transport to move to their destination in the centre.

We instantiate the model presented in section 3 as follows. The utility of a trip *t* is defined based on importance, quality of service and cost, as follows:

$$U(t) = \frac{Im(t) \cdot QoS(t)}{C(t)} \qquad (5)$$

The idea is to define strategies that assign the available parking spaces based on the utility of the trips. For this, we consider three different utility functions (*Baseline*, *Vehicle emission* and *Vehicle emission per person*) as described below.

*Baseline (B).*

Here no prioritization is done, that is all cars have a constant importance of 1 and a constant cost of 1. Thus, the utility of any trip is:

$$U(t) = QoS(t) \qquad (6)$$

Also, for all cars, entering the city implies an increment in QoS by a constant *q*, which is the same for all cars. Thus, the following holds for all trips:

$$U(t_i|I) = U(t_i|\bar{I}) + q \qquad (7)$$

And trivially, (3) is maximized if all requested parking spaces are assigned to any car.

*Vehicle emission (VE).*

In this case, prioritization is done based on the emissions of a car. The importance is the same for all cars (3) and the cost of a trip depends on the average emissions of the vehicle. Thus:

$$U(t) = \frac{QoS(t)}{e(t)} \qquad (8)$$

where $e(t)$ is a measure representing the average emission of the vehicle carrying out the trip *t*. We consider that vehicles belong to different emission types, which are known a priori, and which are used to estimate *e(t)*. With this, it holds:

$$U(t_i|I) = U(t_i|\bar{I}) + \frac{q}{e(t)} \quad (9)$$

Thus, for maximizing (3), and since $q$ is constant for all trips, the assignment should be done by prioritizing trips of vehicles with less emissions.

*Vehicle emission per person (VEP).*

In this case, we consider that vehicles can carry more than one person and the prioritization is done with regard to emissions per person. Thus:

$$U(t) = \frac{n(t) \cdot QoS(t)}{e(t)} \quad (10)$$

where $n(t)$ represents the number of people in a vehicle. That is, the importance of a trip is proportional to the passengers in the car. In this case:

$$U(t_i|I) = U(t_i|\bar{I}) + \frac{n(t) \cdot q}{e(t)} \quad (11)$$

For maximizing (3), the assignment should be done by prioritizing trips of vehicles with a better ratio of emission per person.

One of the problems of prioritization in a dynamic environment like the one we consider here, is the fact that the vehicles request the parking spaces at different times. There is no direct knowledge of upcoming future requests and, thus, prioritization among requesting cars is not straightforward. To overcome this problem, we use the following idea. Let *pa* be the number of existing parking spaces in the city and let *poc* be the parking spaces that are occupied at a given moment. We specify a threshold $\theta_{LP} < pa$ from which on the access to parking spaces should be limited. Then, we define a parking-assignment level $k_p$ as follows:

$$k_p = \begin{cases} 1 \text{ if } poc \leq \theta_{LP} \text{ (no restrictions)} \\ 0 \text{ if } poc \geq pa \text{ (no vehicles allowed)} \\ \frac{(pa - \text{poc})}{(pa - \theta_{Lp})} \text{ otherwise} \end{cases} \quad (12)$$

$k_p$ can be calculated at specific time intervals and represents the ratio of cars that can be assigned to parking spaces. If the occupancy is below the threshold $\theta_{LP}$, a parking space will be assigned to any requesting vehicle. As the occupancy grows towards the capacity, the percentage of vehicles that will be assigned decreases proportionally, up to full occupancy. We can use $k_p$ to prioritize assignments. Parking spaces will be assigned to the ($k_p$·100)% of vehicles with lowest emissions for the *VE* strategy, or to the ($k_p$·100)% of vehicles with lowest emissions per person for the *VEP* strategy during a time interval. In order to implement this prioritization scheme, we assume that all cars belong to different known emission classes and that we can estimate the average percentage of vehicles that belong to each class (e.g., from historical data). Furthermore, we assume that a vehicle that requests access notifies the emission type it belongs to and the number of people it carries. Based on these assumptions, the prioritization can be implemented as follows. We order the existing vehicle types by their average emissions / average emissions per person (up to 5 people) and weight each entry in this order by the expected percentage of appearance. Then, given a vehicle and a value of $k_p$, the vehicle can enter the centre if it belongs to the ($k_p$·100)% of vehicles in this order.

In the case of the baseline strategy, since no prioritization is done, parking spaces are assigned in a first-come-first-served manner up to the full capacity.

*4.1. Experimental evaluation*

We carried out experiments using SUMO[1] [2], a popular open-source microscopic traffic simulator. Among other characteristics, SUMO allows defining traffic infrastructures (including road segments, intersections, parking areas, etc.) and specify vehicles' trips (origin, destination, initial time, etc.). SUMO simulations can be run using a GUI. However, we used TraCI (Traffic Control Interface), which provides access to SUMO core and allows on-line manipulation of objects (e.g. vehicles).

The aim of the experiments is to check how the different strategies perform with regard to the number of cars and people that enter the city as well as the pollutants emitted by vehicles in the centre. For the latter, SUMO includes several emission models. We used HBFA3, which is based on the database HBEFA[2] version 3.1. The model simulates several vehicle emission pollutants, and we chose $NO_x$ as the reference in our experiments.

SUMO implements several vehicle emission classes including heavy duty, passenger, and light delivery emission classes, combined with different EU emission standards (levels 0-6). In the experiments, we chose 6 different types of vehicles with different emission classes: eVehicle ("Zero/default"), gasolineEuroSix ("HBEFA3/LDV_G_EU6"), dieselEuroSix ("HBEFA3/LDV_D_EU6"), hovDieselEuroSix ("HBEFA3/PC_D_EU6"), normalVehicle ("HBEFA3/PC_G_EU4"), and highEmissions ("HBEFA3/PC_G_EU3.

While SUMO is able to provide information about pollutants emitted by vehicles, it does not include a model of how those pollutants evolve in the air. These values depend on many different factors, not only different pollutant emission sources but also on weather conditions (wind, rain, temperature changes, etc.).

Typically, pollution data in cities are measured by atmospheric stations that measure the pollutants in the air, and thus depend on more factors than just direct emissions of vehicles. The basic idea of the air quality model used in our experiments is that pollution at a given time $t$ is the sum of a pollution (due to effects not related to traffic) plus the pollution due to vehicles. With this idea, we define the pollution at time $t$ by:

$$p_t = pc_t + pe_t, \tag{13}$$

where

$$pc_t = 0.7 \cdot pc_{t-1} + 100000 \cdot 0.3 \cdot \lambda c_t \tag{14}$$

and

$$pe_t = max(0; p_{t-1} - pc_t + e_t - 10000 \cdot \lambda e_t) \tag{15}$$

$p_{t-1}$ and $pc_{t-1}$ are the global pollution and the static pollution values at time *t-1*. $e_t$ is the pollution emitted by vehicles between the time interval from t-1 to t. The static pollution is set closed to a constant of 100000 mg NOx and the pollution generated by cars is diminished in each time step due to atmospheric effects by a constant of 10000 (we set these values empirically). $\lambda c_t$ and $\lambda e_t \in$ [0.9, +1.1] are uniformly randomly generated factors that represent possible random changes in the atmosphere. While we recognise that this is a simplification of the real world, this measure of pollution allows us to analyse and compare the different control strategies.

---

[1] https://www.eclipse.org/sumo/

[2] http://www.hbefa.net/

For the experiments we designed a virtual city as shown in Fig. 1. It consists of a 6 x 6 km square with a city centre (control zone; represented in grey) with eight access points. The network is made up of road segments connected by intersections. All segments are bidirectional, and all intersections are roundabouts. As shown in Fig. 1, there are 8 parking spaces in the city centre. Furthermore, there are 4 parking spaces outside the centre, that will be used by cars that cannot enter the centre.

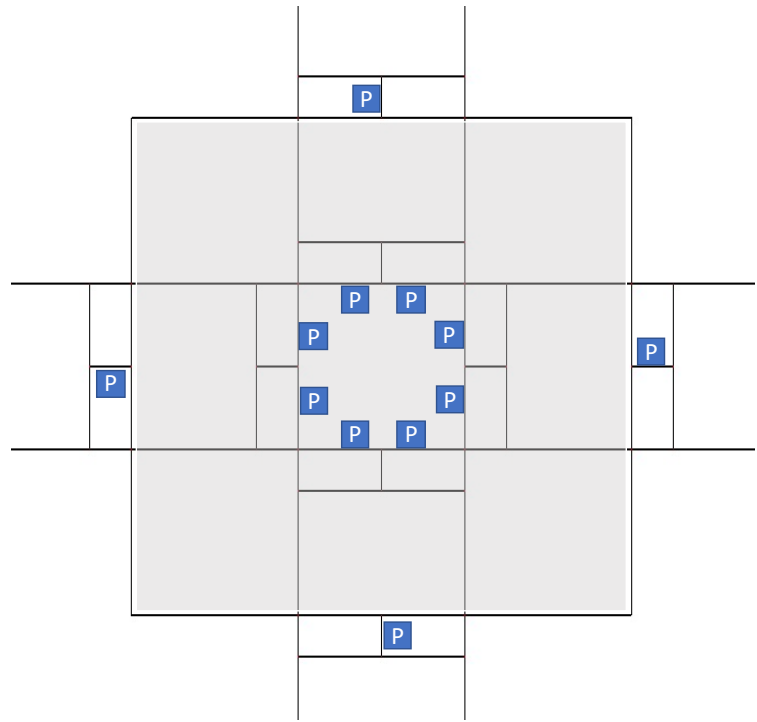

**Figure 1. Schematic** road network used in the experiments: black lines are roads, blue rectangles are parking areas and the grey area represents the city centre, e.g., the control zone for prioritization.

Trips are generated randomly from any outside point to the closed parking area in the city centre. Each trip is carried out by a vehicle (with a specific emission type) and transports between 1 and 5 passengers. Once a vehicle reaches the parking it will park for half an hour and then go back to its origin. Vehicles request to park in the centre when they are getting nearby the central area (between 500 and 100 meters). The control strategy, based on the available parking spaces in the city, decides whether a vehicle can enter or not. Vehicles that cannot enter, reroute to the closest parking outside the centre. They park there for 1 hour (including half an hour additional time for the persons to move to the centre and back) and then move back to their origin.

We ran two simulations with different vehicle arrival rates. One simulation of 5 hours with a fixed rate of 1000 vehicles per hour (generated with an exponential distribution) and another simulation of 5 hours with varying arrival rates (1000 vehicles per hour during the first 2 hours, 2000 vehicles the third, 3000 vehicles the fourth and 2000 the fifth hour).

Origin, number of people per vehicle (between 1 and 5), and vehicle emission types are chosen randomly. All 5 vehicle emission types have the same probability, except *high-Emissions*, which has half the rate of the others.

The time interval to update $k_p$ and the pollution $p_t$ is set to 60 seconds. The available parking capacity in the city is set to 550 and $\theta_{LP}$=500. The capacity of parking spaces outside the centre is unlimited.

*4.2. Results*

Table 1 shows a summary of the results of the experiments for the simulations with fixed arrival rate. For each strategy we present the number of vehicles (out of a total of 5079 vehicles) that entered the centre because they got a parking space, the number of persons who could enter the centre (out of a total of 15353), the total amount of $NO_x$ (mg) emitted by vehicles in the centre, the average time per trip and the average time per person (from origin to the centre and back).

**Table 1.** Simulation results parking access with constant arrival rates.

| Strategy | #vehicles with access | #persons with access | NOx emitted in centre | Avg. time per trip (s) | Avg. time per person (s) |
|---|---|---|---|---|---|
| Baseline | 4770 / 93.9% | 14414 / 93.9% | 3750537 | 2771 | 2771 |
| VE | 4477 / 88.1% | 13538 / 88.2% | 2962915 | 2852 | 2852 |
| VEP | 4482 /88.2% | 14004 / 91.2% | 3045778 | 2849 | 2807 |

The results show that the highest number of vehicles and people can enter the city with the baseline strategy. This is because this strategy obtains the highest occupation of parking slots. With the two prioritization strategies, the obtained occupation is a bit lower, due to the fact that only a percentage of cars are allowed to enter the city when the parking occupation reaches $\theta_{LP}$. Both methods, VE and VEP, have similar occupation ratio and thus, about the same number of cars are allowed to enter the centre. With regard to the average time per vehicle and per person, also the baseline strategy has the best performance, again, because more vehicles are allowed to enter the centre and thus, spend less time in the system (as we mentioned before, parking outside, takes half an hour longer). Regarding the number of people that gain access, there is a clear improvement from VE to VEP, since VEP prioritizes access not only with regards to emissions, but also the number of people in a vehicle. Considering the NOx emissions, VE and VEP clearly outperform the baseline strategy considerable. This can also be observed in Figure 2, which shows the evolution of the pollution in the air ($p_t$), over time.

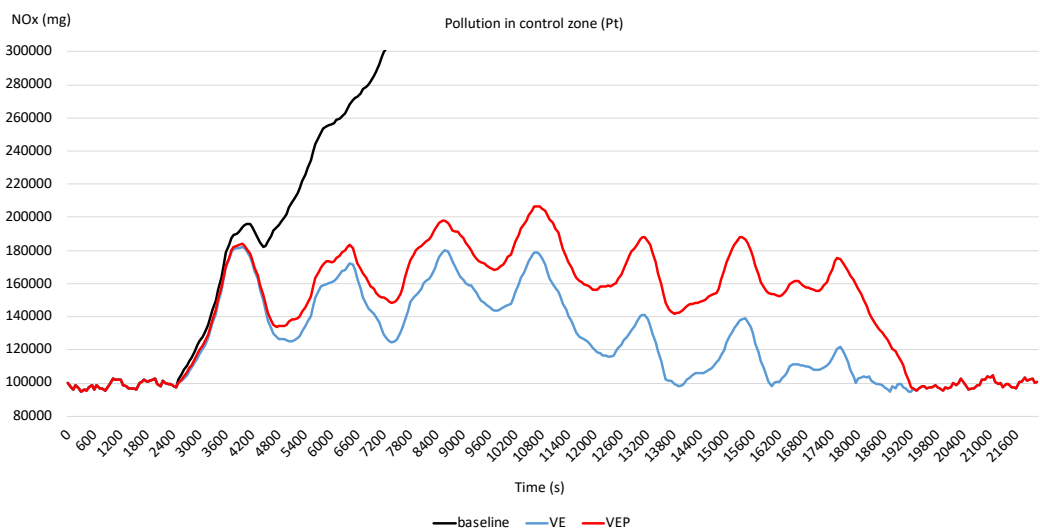

**Figure 2.** Evolution of the pollution ($NO_x$) in the control zone during the simulation for parking access with constant arrival rates.

The pollution obtained with the baseline strategy grows very fast and exceeds at its worst point $7 \cdot 10^5$. In contrast, both other strategies, that prioritize the assignment of parking spaces, and thus, of the access to the city centre, for vehicles with lower emissions maintain the pollution acceptable and fairly stable values. The oscillations observed in VE and VEP curves reflect the inertia of the system to react to control actions (allowing higher or lower rate of vehicles to enter the centre). That is, when access is reduced (i.e., parking-assignment level $k_p < 1$) vehicles in the centre keep on producing NOx. Therefore, it takes some time until outflow of vehicles from city centre surpasses inflow thus producing an increase of $k_p$.

The advantages of our control strategies become more apparent when the arrival rates of vehicles change with time. Table 2 and Figure 3 show the results of the experiment with such varying arrival rates of vehicles. In these experiments, the total number of vehicles is 9030 with a total of 27272 people.

**Table 2.** Simulation results parking access with varying arrival rates.

| Strategy | #vehicles with access | #persons with access | NOx emitted in centre | Avg. time per trip (s) | Avg. time per person (s) |
| --- | --- | --- | --- | --- | --- |
| Baseline | 4837/53.6% | 14458/53.0% | 3833958 | 3366 | 3374 |
| VE | 4639/51.4% | 14015/51.4% | 1894476 | 3394 | 3394 |
| VEP | 4644/51.4% | 15348/56.3% | 2044755 | 3394 | 3324 |

In this case, again the highest number of vehicles that can enter the city centre and the lowest average time per trip are achieved with the baseline strategy. However, when considering persons, the best values are obtained with the VEP strategy. This is because after 2 hours there are more vehicles that want to enter the centre and thus, any prioritization strategy will find "better" vehicles to assign the parking spaces. This is also the reason why the pollution drops considerably for the VE and VEP strategies at about 9000 seconds in the experiments, as it can be observed in Figure 3. The pollution values for the baseline strategy here grows to about $8*10^5$. Also considering the NOx emissions in the centre, a clear and considerable improvement can be observed for the prioritization strategies.

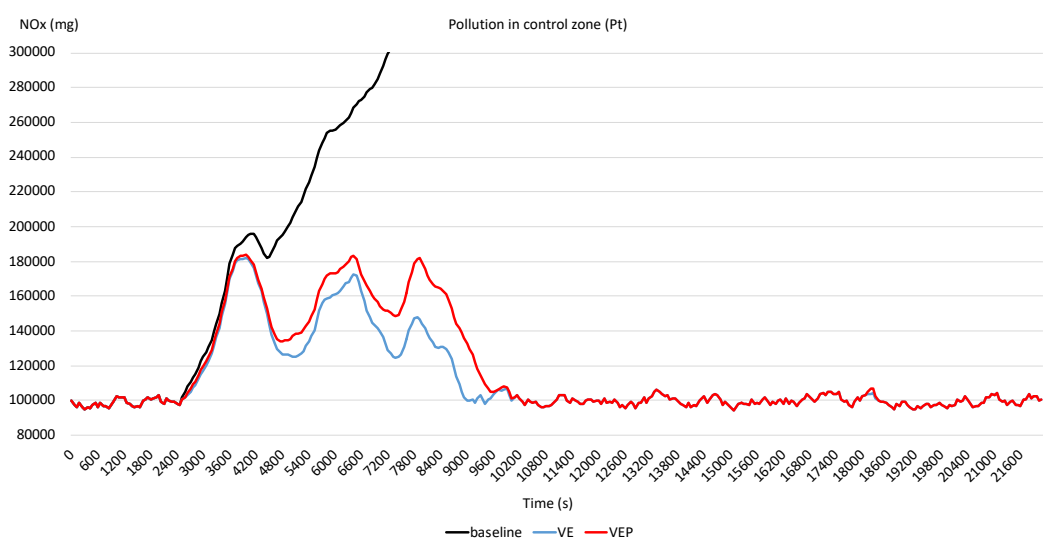

**Figure 3.** Evolution of the pollution (NOx) in the control zone during the simulation for parking access with varying arrival rates.

## 5. Use Case: pollution-based access to city centre

In the second use case, again we assume that people want to use their cars to enter the centre of a city in order to accomplish some tasks, but instead of restricting the access because of limited parking spaces, we want to control the access to the centre based on the current pollution values. The idea is to specify regulation devices at all entry points of a specific sensible area of a city (e.g., the city centre) that track the current pollution and dynamically grant or restrict the access of vehicles based on admissible pollution values. In this process, the devices prioritize the access of vehicles that have a higher importance from a social point of view.

We analyse different control strategies and consider that the system implementing a control strategy works as follows. A vehicle that wants to enter the restricted area, requests access at an entry point and the control strategy either grants or denies this access. Vehicles that are allowed to enter can park in the centre and vehicles that are not allowed to enter have to find parking space outside and their drivers may take public transport to move to the centre.

The idea of the strategies is to restrict access to the centre in such a way that the measured pollution in the area ($p_t$) is kept below a certain maximum at any time *t*. That means that, in contrast to the parking spaces, here the number of available resources is not a fixed number but may change with the measured pollution.

The following idea is applied. We define three thresholds $\theta_N$, $\theta_L$ and $\theta_H$. $\theta_H$ represents the maximum allowed pollution value that should not be exceeded. $\theta_L < \theta_H$ is the level of pollutions from which on restrictions are applied and $\theta_N < \theta_L$ is the level of pollution that is considered normal. Restrictions are activated when $p_t > \theta_L$ and until $p_t < \theta_N$. If restrictions are applied, only $k_e$ percent of the vehicles can enter the centre. However, if $p_t > \theta_H$ access to the centre is denied for any vehicle.

When restrictions are active at a given time *t*, we calculate the number of vehicles $v_t$ that are allowed to enter in the following time interval by applying the ideas of a proportional derivative (PD) controller:

$$v_t = vc_{t-n} \cdot \left( cp \cdot \frac{(\theta_H - p_t)}{(\theta_H - \theta_L)} + cd \cdot \frac{(p_{t-1} - p_t)}{(\theta_H - \theta_L)} \right) \quad (16)$$

where $vc_{t-n}$ are is the number of vehicles that were circulating (not parking) in the centre in the time interval *t-n*, $p_t$ is the pollution at the beginning of time interval *t* and $p_{t-1}$ the pollution at the beginning of the previous time interval. *cp* and *cd* are constants that are applied for the deviation of the current $p_t$ from the maximum allowed value (first factor) and for the proportional increment of $p_t$ with respect to its previous value, respectively. That is, the new allowed number of vehicles to enter the city depends on the vehicles that were circulating at a previous moment (*t-n*), the increment of the pollution (derivative component of PD controller) and how far the pollution value is from the maximum (proportional component of PD controller). We consider $vc_{t-n}$ because the effect that the vehicles entering the centre have on the pollution is observed later and it depends on the time the vehicle will actually stay in the centre.

Once we have calculated $v_t$ we need a mean to grant access by prioritizing certain trips. Here again we use the idea employed in section 4. We translate the allowed vehicles to an access-level $k_e$ that represents the ratio of vehicles that can enter the centre. In particular, we determine $k_e$ in relation to the received access requests in the previous time interval:

$$k_e = \min\left(1, \frac{v_t}{\max(1, ve_{t-1})}\right) \quad (17)$$

where $ve_{t-1}$ is the number of vehicles that requested access in the previous time interval.

In order to maximize (3), the prioritization is accomplished by granting access to the ($k_e \cdot 100$)% vehicles with highest utility gain.

Similar to section 4 and based on the three different utility functions specified there, we define the following control strategies:

- *Baseline* – all vehicles have the same priority
- *Vehicle emission (VE)* – vehicles with lower emissions have a higher priority
- *Vehicle emission per person (VEP)* – vehicles with a lower ratio of emissions per person have a higher priority.

The prioritization scheme is implemented in the same way as described in section 4. That is, the existing vehicle types are ordered by their average emissions / average emissions per person, each entry in this order is weighted by the expected percentage of appearance, and a vehicle is granted the access to the centre if it belongs to the ($k_e$·100)% of vehicles in the corresponding order. In the case of the baseline strategy, the ($k_e$·100)% vehicles are randomly chosen.

In addition to the aforementioned strategies, we define two other strategies:

*Ratio Reduction Emission (RRE).*

In this case, the access value $k_e$ is not applied to the ratio of vehicles that can enter the restricted area. Instead, $k_e$ represents the ratio of emissions that are allowed to be generated with respect to the normally generated emissions in the same moment or time frame. Given $k_e$, we calculate the ratio $k_e'$ of vehicles with lowest emissions (with respect to the normal demand) that together produce the ($k_e$*100)% of the emissions usually generated in the same moment or time frame. It holds that $k_e' \geq k_e$. Afterwards, the strategy applies the same prioritization scheme as *VE*.

*Ratio Reduction Emission per Package (RREP).*

As the RRE strategy, here $k_e$ is translated to a ratio of vehicles $k_e'$. Then, the same prioritization scheme as in VEP is employed with the new ratio.

*5.1. Experimental evaluation*

We used the same experimental setup as described in section 4. The time interval to update $k_e$ and the pollution $p_t$ is set to 60 seconds. The values of the other parameters are set as follows: $\theta_N = 110000$, $\theta_L = 200000$, $\theta_H = 300000$, $n = 2100$ (for determining $vc_{t-n}$), $cp = 0.1$, $cd = 2$. The values for *n, cp* and *cd,* have been chosen empirically.

Again, we ran the simulations with the two different vehicle arrival rates (5 hours with a fixed rate of 1000 vehicles per hour and 5 hours with 1000, 1000, 2000, 3000 and 2000 vehicles per hour). In the first experiments 5079 vehicles request access (with a total number of 15353 people). In the second experiments the total number of vehicles is 9030 with 27272 people.

Table 3 and Figure 4 show a summary of the results of the experiments with the 5 different control strategies applied to the case of the constant arrival rate. For comparison, we also added the results that are obtained if no control strategy is applied (*no control*). This strategy shows the impact of applying control strategies on the change in the pollution.

**Table 3.** Simulation results pollution-based access with constant arrival rates.

| Strategy | #vehicles with access | #persons with access | NOx emitted in centre | Avg. time per trip (s) | Avg. time per person (s) |
|---|---|---|---|---|---|
| no control | 5079/100% | 15353/100% | 3996314 | 2684 | 2684 |
| baseline | 4063/ 80% | 12258/79.8% | 3136958 | 2963 | 2965 |
| VE | 4534/89,3% | 13765/89.7% | 3145401 | 2835 | 2830 |
| RRE | 4620/91% | 13953/90.1% | 3151273 | 2812 | 2814 |
| VEP | 4542/89.4% | 14078/91.7% | 3153845 | 2834 | 2802 |
| RREP | 4581/90.2% | 14346/93.4% | 3151093 | 2824 | 2779 |

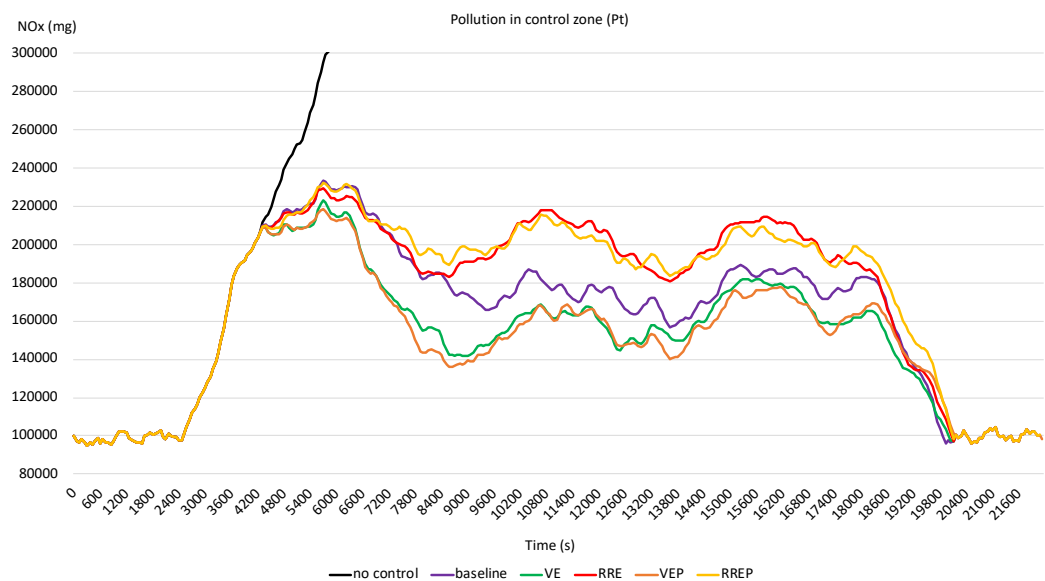

**Figure 4.** Evolution of the pollution ($NO_x$) in the control zone during the simulation for pollution-based access with constant arrival rates.

As it is obvious, all cars can enter the centre if no control strategy is applied and also the best average times per vehicle and per person are obtained. However, this case does not control the emissions and very fast, the pollution in the control zone exceeds the maximum allowed pollution of $3·10^5$ mg NOx. At the worst point in the simulation (not shown in the figure), the pollution value almost reaches $10^6$ mg NOx.

Regarding the control strategies, all are able to keep the pollution below the maximum allowed level. As expected, the *baseline* strategy restricts the access to more vehicles than the others, since it randomly chooses, which vehicles are allowed to access. *VE* and *VEP*, obtain similar results granting access to much more vehicles than *baseline*. These strategies' approach is to reduce the same percentage of vehicles as *baseline* but they select less contaminant vehicles. The effect is that less contaminants are released, thus pollution is reduced and, consequently, more vehicles are allowed to enter the control zone. With respect to *RRE* and *RREP*, the pollution values are higher, but still below the maximum allowed level. These are the strategies that allow more vehicles to enter the control zone. Their approach is to allow access to less contaminant vehicles that jointly add up a percentage of expected emissions. That is, they restrict access to fewer but high contaminant vehicles. This effect, however, is mitigated as the simulation advances, because the control system adapts the vehicles that can enter the centre to the value of the actually measured pollution values. Still, there is roughly a 2% of additional vehicles that can enter. As shown in table 3, the number of people that can enter increases about 2-3% if prioritization also takes this number into account (as in the strategies VEP and RREP versus VE and RRE).

As it can be noted in Figure 4, triggering limitations (at a pollution level of 200000) does not have an immediate effect on reducing $p_t$. This is also the reason why $p_t$ oscillates roughly between 140000 and 230000.

The results for the experiments with varying arrival rates of vehicles (9030 vehicles with a total of 27272 persons) are presented in Table 4 and Figure 5.

**Table 4.** Simulation results pollution-based access with varying arrival rates.

| Strategy | #vehicles with access | #persons with access | NOx emitted in centre | Avg. time per trip (s) | Avg. time per person (s) |
|---|---|---|---|---|---|
| no control | 9030/100% | 27272/100% | 7862495 | 2735 | 2736 |
| baseline | 4096/45.4% | 12398/45.5% | 3156139 | 3479 | 3478 |
| VE | 5809/64.3% | 17683/64.8% | 3136873 | 3215 | 3208 |
| RRE | 6445/71.4% | 19505/71.5% | 3266085 | 3112 | 3110 |
| VEP | 5739/63.6% | 18515/67.9% | 3193441 | 3228 | 3167 |
| RREP | 6180/68.4% | 20005/73.4% | 3290520 | 3157 | 3088 |

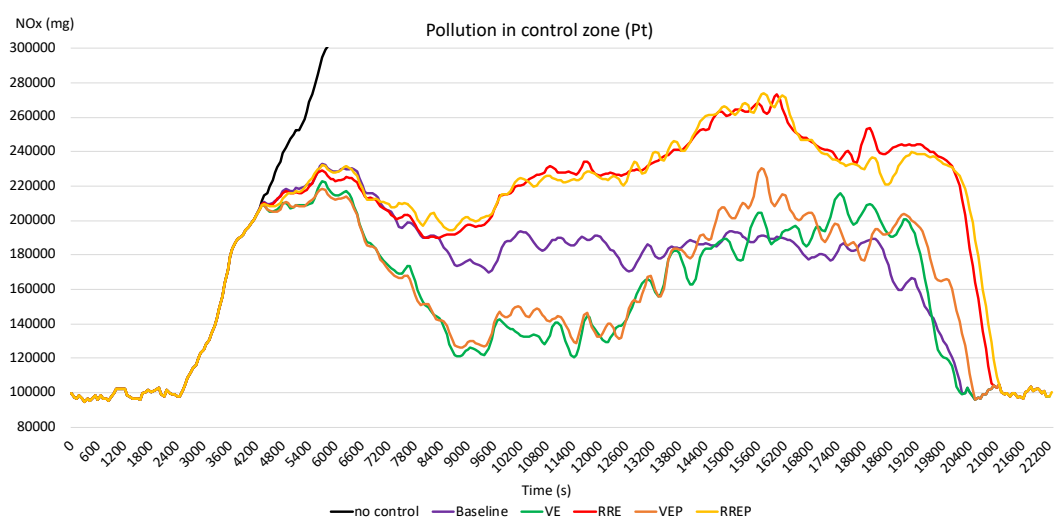

**Figure 5.** Evolution of the pollution (NOx) in the control zone during the simulation for parking access with varying arrival rates.

In general, the results show similar conclusions to the ones obtained with constant arrival rates. When no control is applied, the pollution increases very fast and exceeds the allowed maximum. The other strategies keep the pollution below that maximum. Again, *VE* and *RRE* allow more vehicles to enter than baseline, with acceptable pollution values. In this case, with a higher demand, the difference is much more significant. *VEP* and *RREP* obtain similar results but allow more people to enter. Average trip times are in line with the rate of vehicles that could/could not access the control zone. Strategies that take into account the number of persons in the vehicles benefited of slightly lower travel times per person as compared to just considering emissions.

It should be noted that the presented evaluation experiments are based on schematic simulations and are meant to provide a general outlook on the expected performance of the proposed solution. We did not use a scenario of a particular location or city. However, the solutions could be applied to many urban areas. Furthermore, we are aware that we did not consider certain issues that may occur in real world scenarios and could influence the systems performance. For instance, in our approach, we rely on information provided or obtained from the cars (number of people and emission type). Here, we assume the number of people reported by users (e.g. via an app or vehicle-to-infrastructure communication) is correct. There are means to avoid malicious behaviour. For example, in the city of Madrid (Spain), traffic authorities use cameras and penalties to avoid unauthorized use of priority lanes with a restriction on the occupants of a car. In the future, smart cars could detect the number of people and communicate that information to the infrastructure. The type of car could easily be identified reading license plates. Also, we have not explicitly dealt with hybrid vehicles which may run either with combustion or with electric engines while in the city centre. A simple option for incorporating them into the proposed solution consists in using their average emission pattern. If their battery-level could be elicited in a trusted manner, it could be used to estimate their expected pollution level even better. In any case, our approach is based on dynamic control and erroneous emission predictions or other contingencies that would result in higher pollution values, would imply increasing the access limitations for the next time intervals. Thus, the system would implicitly adapt to such cases. This holds also for certain public services for which restrictions are not applied (e.g., ambulances, police, buses, etc.).

## 6. Conclusions

Organising urban mobility in large cities is one of the big challenges that governments must face nowadays. The demand for mobility services, both of people and goods, requires an efficient usage of the transportation infrastructure, whose capacity is frequently exceeded by the demand. This causes frustrations in individuals (e.g. delivery delays, time spent in cars, extra fuel consumption, …) and also to the society in general (e.g. noise, pollution, etc.).

In this paper, we have argued that the use of infrastructure elements should be regulated and prioritized in a way that benefits the social or global utility. In particular, we present a general assignment model that allocates limited transportation resources to the traffic activities (trips) that have a higher utility from the point of view of the society. Here utility is considered to have three components: i) "importance" of the transportation activity, ii) quality of service, and iii) cost from the society's point of view (e.g. emissions).

We have instantiated the model in two different use cases. In the first one, the limited transportation elements are the available parking places in a city centre. Vehicles are allowed to enter the centre if they first obtain a parking reservation, which is granted to vehicles according to their global utility when the availability of parking is below some threshold. We proposed three strategies: (i) all vehicles have the same priority, (ii) vehicles with lower emission rates have higher priority, and (iii) priority depends on a combination of emissions and the "importance" of the trip (in terms of number of people travelling). In the second use case, we present a control system that dynamically determines the access limitation level to a city centre based on the current measures of environmental pollution. Again, the system employs a prioritization strategy that determines which vehicles (trips) can enter the area and which cannot. We have used the same strategies as in the first use case and added two additional ones that focus on reducing the expected emissions rather than the expected number of vehicles.

We have carried out several experiments with the traffic simulation tool SUMO to analyse the performance of our proposal with different strategies. As a conclusion we can determine that the general idea of dynamically limiting the access to a restricted area allows to maintain the environmental pollution in this area below given limits. Furthermore, a prioritization of access based on emissions and/or "importance" of a trip improves the utility of the system and allows to accomplish in an efficient way more of the important transportation tasks under the given pollution limits.

Given the prioritization methods, less important tasks or the use of vehicles with higher emissions will imply more restrictions and mobility limitations. As a side effect, users may tend to acquire more environment-friendly vehicles and may try to collocate different transportation tasks in single movements. In this way, they could benefit from higher priorities in the use of the infrastructure.

The presented use cases have been treated from the point of view of current existing infrastructures and existing types of vehicles. In this regard, in the future, changes can be expected especially with the appearance of autonomous vehicles. Autonomous vehicles, for example, may not suffer the problem of limited parking spaces in city centres, since they could search parking spaces in external areas after leaving the occupants in the centre [32]. Nevertheless, the second use case may still apply, because access restrictions for cars to city centres may be an issue also in the future, either because of high pollutions or for other reasons such as traffic congestions.

With regard to future lines of research, regarding the use cases, it would be interesting to carry more realistic simulations (e.g., with a real city and real traffic data). Nevertheless, we do not think that the comparison results would change considerably. Furthermore, we plan to extend our model for prioritising other than time-unlimited infrastructure elements. Another line or research is the definition of "importance" of different types of trips, probably using semantic technologies.

**Author Contributions:** Conceptualization, H.B., A.F., P.M., J.P.T. and S.O.; methodology, H.B., A.F., P.M., J.P.T. and S.O.; software, H.B., A.F; validation, H.B., A.F. and S.O.; formal analysis, H.B., A.F., P.M., J.P.T. and S.O.; investigation, H.B., A.F., P.M., J.P.T. and S.O.; writing—original draft preparation, H.B., A.F. and S.O.; writing—review and editing, H.B., A.F., P.M., J.P.T. and S.O.; visualization, H.B., A.F. and S.O.; funding acquisition, H.B., A.F. and S.O. All authors have read and agreed to the published version of the manuscript.

**Funding:** This work has been partially supported by the Spanish Ministry of Science, Innovation, and Universities, co-funded by EU FEDER Funds, through grant RTI2018-095390-B-C31/32/33 (MCIU/AEI/ FEDER, UE).

**Conflicts of Interest:** The authors declare no conflict of interest.